\documentclass[
]{ceurart}

\usepackage{listings}
\usepackage{xcolor}
\usepackage{makecell}
\usepackage{hyperref}
\begin{document}

\pagestyle{myheadings}

\copyrightyear{2023}
\copyrightclause{Copyright for this paper by its authors.
  Use permitted under Creative Commons License Attribution 4.0
  International (CC BY 4.0).}

\conference{Ontology Showcase and Demonstrations Track, 9th Joint Ontology Workshops (JOWO 2023), co-located with FOIS 2023, 19-20 July, 2023, Sherbrooke, Québec, Canada.}

\title{OPPO: An Ontology for Describing Fine-Grained Data Practices in Privacy Policies of Online Social Networks}

\author{Sanonda Datta Gupta}[%
email=sanonda.gupta@maine.edu
]

\cormark[1]

\author{Torsten Hahmann}[%
email=torsten.hahmann@maine.edu
]

\address
{School of Computing and Information Science, University of Maine, ME, USA}

\cortext[1]{Corresponding author.}

\begin{abstract}
Privacy policies outline the data practices of Online Social Networks (OSN) to comply with privacy regulations such as the EU-GDPR and CCPA. Several ontologies for modeling privacy regulations, policies, and compliance have emerged in recent years. However, they are limited in various ways: (1) they specifically model what is required of privacy policies according to one specific privacy regulations --  GDPR; (2) they provide taxonomies of concepts but are not sufficiently axiomatized to afford automated reasoning with them; and (3) they do not model data practices of privacy policies in sufficient detail to allow assessing the transparency of policies. This paper presents an OWL \textbf{O}ntology for \textbf{P}rivacy \textbf{P}olicies of \textbf{O}SNs -- OPPO -- that aims to fill these gaps by formalizing detailed data practices from OSNS' privacy policies. OPPO is grounded in BFO, IAO, OMRSE, and OBI, and its design is guided by the use case of representing and reasoning over the content of OSNs' privacy policies and evaluating policies' transparency in a greater detailed.
\end{abstract}


\maketitle

\section{Introduction}
\label{sec:Intro}
The widespread use of Online Social Network (OSN) raises many privacy concerns, such as storing users' personally identifiable information (PII) longer than required or without following proper security mechanisms. To mitigate such concerns, various privacy regulations (eg., GDPR \cite{GDPR} and CCPA\cite{CCPA}) have been introduced, which require organizations to be transparent about their data practices (collection, processing, or storage of users' data). An essential aspect to comply with these regulations are the privacy policies that describe the data practices of OSNs. However, privacy policies are usually long and complex and often do not explain the data practices in sufficient detail \cite{cate2010limits,mcdonald2008cost}. For instance, a policy may mention that it follows best practices in security without actually specifying what mechanisms it uses and to which types of data (personal vs. non-personal) they apply. Such generic or overly vague descriptions are indicators of an OSN's \textit{lack of transparency} in informing the users of its \textit{actual} data practices. In this paper, we present the \textbf{O}ntology for \textbf{P}rivacy \textbf{P}olicies of \textbf{O}SNs (OPPO) that is designed to encode the data practices from the privacy policies of OSNs (and other companies) in as much detail as possible.  The ontology is intended to let the OSN, users, and others (e.g.\ privacy advocacy groups) query the formal ontological representations of OSNs' data practices for specific details or the general level of detail they provide. 

Prior research on formal representations of online privacy has focused on extracting rights and obligations to check compliance with regulations \cite{breaux2006towards,kiyavitskaya2008automating,yue2011systematic} and  developed privacy related vocabularies and taxonomies \cite{palmirani2018pronto,pandit2019creating,oltramari2018privonto,lioudakis2021gdpr,fatema2017compliance}. However, most of these works miss more detailed yet important concepts and relations, such as response time for data rectification or erasure requests, retention duration (limited or unlimited) for different data types (personal vs.\ non-personal), or security mechanisms that are applied to different storage entities (device vs.\ data center). These details are not omitted just for a lack of methods that are able to extract such details but, more fundamentally, the ontologies (e.g., Pronto \cite{palmirani2018pronto} and PrivOnto \cite{oltramari2018privonto}) lack important concepts for modeling data practices at such levels of  details in the first place. 

Towards the goal of filling this gap, we develop OPPO as an upper-level ontological model for the domain of privacy policies and data practices. It is grounded in top-level ontological distinctions and capable of describing data practices at more granular levels. 
As a proof-of-concept of OPPO's ability to describe fine-grained data practices, we focus on introducing classes and properties for capturing data retention, storage, and security practices in detail, including aspects such as how users can request data rectification or what types of security mechanisms are applied to various data types. OPPO is developed following the methodology METHONTOLOGY \cite{fernandez1997methontology} and is guided by a set of 45 competency questions, manual analysis of the GDPR (Art. 5, 13--17, and 32), CCPA (Art. 1798.100, 1798.125, 1798.135, 1798.140, 1798.81.5), and the privacy policies of ten major OSNs\footnote{WhatsApp \cite{WhatsApp}, Signal \cite{Signal}, Telegram \cite{Telegram}, Twitter \cite{Twitter}, Tiktok \cite{TikTok}, WeChat \cite{WeChat}, SnapChat \cite{SnapChat}, Reddit \cite{Reddit}, Pinterest \cite{Pinterest}, and Instagram \cite{Instagram}}. 
OPPO is grounded in the Basic Formal Ontology (BFO) \cite{arp2015building} as well as the Information Artifacts Ontology (IAO) \cite{smith2013iao}, the Ontology for Biomedical Investigations (OBI) \cite{bandrowski2016ontology}, and the Ontology for Medically Related Social Entities (OMRSE) \cite{hicks2016ontology} as widely-used specializations of BFO. OPPO further reuses parts of the Data Privacy Vocabulary (DPV) \cite{pandit2019creating} and the OWL Time ontology \cite{OWLTime}. 
The ontology is shared via GitHub at \url{https://github.com/SanondaDattaGupta/OPPO-Ontology}; it currently contains  60 classes and relations and is formalized in OWL2 \cite{motik2012ontology}\footnote{\texttt{OPPO\_Ontology\_Full\_Import.ttl} fully imports the ontologies BFO, IAO, OBI, OMRSE, DPV and OWL-Time from which concepts are reused or extended. Because only a small fraction of these concepts are relevant to OPPO, we also provide a version of OPPO that imports only small subsets of IAO, OBI, OMRSE and DPV that are limited to the classes (and their superclasses) and relations that are reused or refined. This version is called \texttt{OPPO\_Ontology\_Minimal\_Import.ttl} and gives a better impression of the concepts introduced by OPPO when loaded into an ontology editor like Protégé.}. To demonstrate its use and to verify its consistency, we populate the ontology with data from the privacy policy of Telegram as an example. Its logical consistency has been verified using the OWL Reasoner Hermit \cite{glimm2014hermit}. It has been validated on a subset of the guiding competency questions, expressing and executing them as SPARQL queries over the Telegram data. 

\clearpage
\section{Use Case}
\label{Sec:Use_Case}

OPPO is designed as a tool for OSN users, privacy researchers, policy makers, regulatory bodies, and organizations to help analyze how OSNs' privacy practices relate to different privacy regulations, identify inconsistencies between practices, or better judge their transparency -- just to name a few examples. Generally, OPPO intends to serve as a tool to formulate and answer questions about the data practices described in privacy policies and, thus, achieve greater transparency while dealing with often long and complex policies. For the first released version of OPPO described here, we focus on capturing the practices that describe storage and retention of data, including the security mechanisms an OSN may employ. 

To guide OPPO's development, we have defined 45 competency questions (CQ) \cite{gruninger1995role}\footnote{The complete set of competency questions are available from the GitHub repository.}. The competency questions help define what terms should be included in the ontology and, later on, they serve as questions to measure whether the ontology can express and answer the questions. These questions include simple queries such as \textit{Where does the OSN store my information?} to more complex questions such as \textit{Which of my personal information will still be available after I delete my account?} Note that our competency questions are not limited to questions about data storage and retention practices but also cover other data practices such as collection or notification mechanisms as well. However, in this paper, we focus on the 27 CQs that pertain to data storage, retention, and security practices. In the following, we present three of those competency questions to illustrate their coverage and the kind of answers we would expect.

\begin{enumerate}\itemsep 2pt
 \item \textit{ Where does the social media site/company store my message?}\\
 An OSN may store users' personal information in several places which differ both in the type of storage (on a device vs.\ a data center) and the storage location (Europe vs.\ United States). Hence, this CQ enables us to capture the location and physical devices where an OSN may store users' personal information.

\item \textit{What contents may be stored for a maximum of 12 months?}\\
Both GDPR and CCPA require organizations to explicitly mention the retention period of the collected personal information. While analyzing the ten privacy policies, we found that some privacy policies (such as Telegram), specify different duration description (such as maximum of X months or as long as they need) for different information types. Thus, this CQ can help us capture and query such details about retention practices and also identify where OSNs lack specificity in such details. 

\item \textit{What types of security mechanism are applied to my photos and private chats?}\\
Privacy regulations require organizations to describe how they ensure the security of the retained information. While most organizations \textit{briefly} describe the security mechanisms, some organizations (such as Signal, WhatsApp, and Telegram) do explicitly mention different security mechanisms (hashing mechanism vs. encryption mechanism) that they apply to different information types (public chat vs. media). Moreover, certain information types are more sensitive (such as bio-metric or health information), and thus may pose higher privacy risks, if not stored following sufficiently secure mechanisms. Hence, this CQ exemplifies the kind of competency questions that help evaluate how transparent and concrete OSNs are about their security and data practices.

\end{enumerate}

\section{Related Work}
\label{sec:relatedwork}

In recent years, several vocabularies, ontologies, and conceptual models for modeling privacy policies, regulations, and compliance have been introduced \cite{jafta2020ontology,esteves2022analysis,kurteva2021consent,lioudakis2021gdpr,kirrane2018scalable} but are primarily concerned with modeling rights and compliance issues with respect to regulations (e.g., BPR4GDPR \cite{lioudakis2021gdpr} and SPECIAL \cite{kirrane2018scalable}) and permissions and prohibitions (e.g., CDMM \cite{fatema2017compliance}, ODRL \cite{ianella2007open}). 
Most closely related to our work are PrivOnto \cite{oltramari2018privonto}, PrOnto \cite{palmirani2018pronto}, and DPV \cite{pandit2019creating}, which are vocabularies and conceptual models that also cover -- to some extent -- data practices. However, they do not capture more granular aspects of data practices such as the duration for which data is stored, the location it is stored in, the specific security protocols that are employed, or the specific process to request to rectify user data. While outlining such detailed practices is not explicitly required by most existing privacy regulations, it can help organizations be \textit{more transparent}.  
In the following, we briefly describe the differences in scope, representation, and other limitations of PrivOnto, PrOnto and DPV that distinguish them from OPPO. 

The purpose of PrivOnto \cite{oltramari2018privonto} is quite different from OPPO: it is designed to annotate paragraphs from policies with concepts from their vocabulary as a kind of semantic tagging but PrivOnto does not allow encoding the content of the policies in a formal representation that it can be automatically reasoned with later on. Moreover, the development of PrivOnto predates the release of GDPR and other privacy regulations; hence the concepts do not tie well to data practices mentioned in regulations. 

PrOnto \cite{palmirani2018pronto} overlaps with OPPO in aspects such as the modeling of different privacy-related data types but lacks a full coverage of types of personal data, such as financial, identity or activity data that we add. A more fundamental difference between PrOnto and OPPO is that Pronto is encoded in the LegalRuleML \cite{athan2015legalruleml} language, a defeasible logic, which restricts the ability to automatically reason with PrOnto\footnote{Reasoning with OWL2 ontologies is supported by a wide range of off-the-shelf reasoners but we are not aware of any tools that can reason directly with LegalRuleML. Instead, specifications in LegalRuleML first need to be converted to another format to enable reasoning as described in \cite{lam_hashmi_2019}.} or to integrate it with other non-legal ontologies. 

DPV \cite{pandit2019creating} provides a comprehensive set of privacy-related terms, including for different data types, purposes, legal entities, and data processing operations (e.g. transmit or store) and encodes it in OWL2. However, it still is more of a taxonomy that does not describe or restrict how these concepts are related to one another axiomatically. Moreover, DPV does not tie the concepts to top-level concepts and does not use ontological analysis tools to structure its taxonomies. For example, while DPV loosely defines a group of \emph{legal roles}, data subjects or regulation authorities are not treated as such. Likewise, multiple distinctions between different types of personal data are made, but these distinctions are not integrated into a single coherent taxonomy. However, we reuse some of DPV's terms where appropriate. But OPPO still introduces an additional  60 concepts such as specific security mechanisms (encryption and hashing), different data types (personal and non-personal), duration descriptions (definite and indefinite) and location descriptions (storage type and spatial location). OPPO further introduces concepts to model specific practices, such as for describing how an OSN deals with requests to rectify or delete data, and the associated properties (e.g., request and response type and response delay). These allow capturing data practices of OSNs more fully and in greater detail, thus enabling querying and evaluating the transparency and level of detail across OSNs.

\section{The Conceptual Model of OPPO}
\label{sec:OPPO}

\begin{figure}[t]
    \centering
   \includegraphics[width=1.0\textwidth]{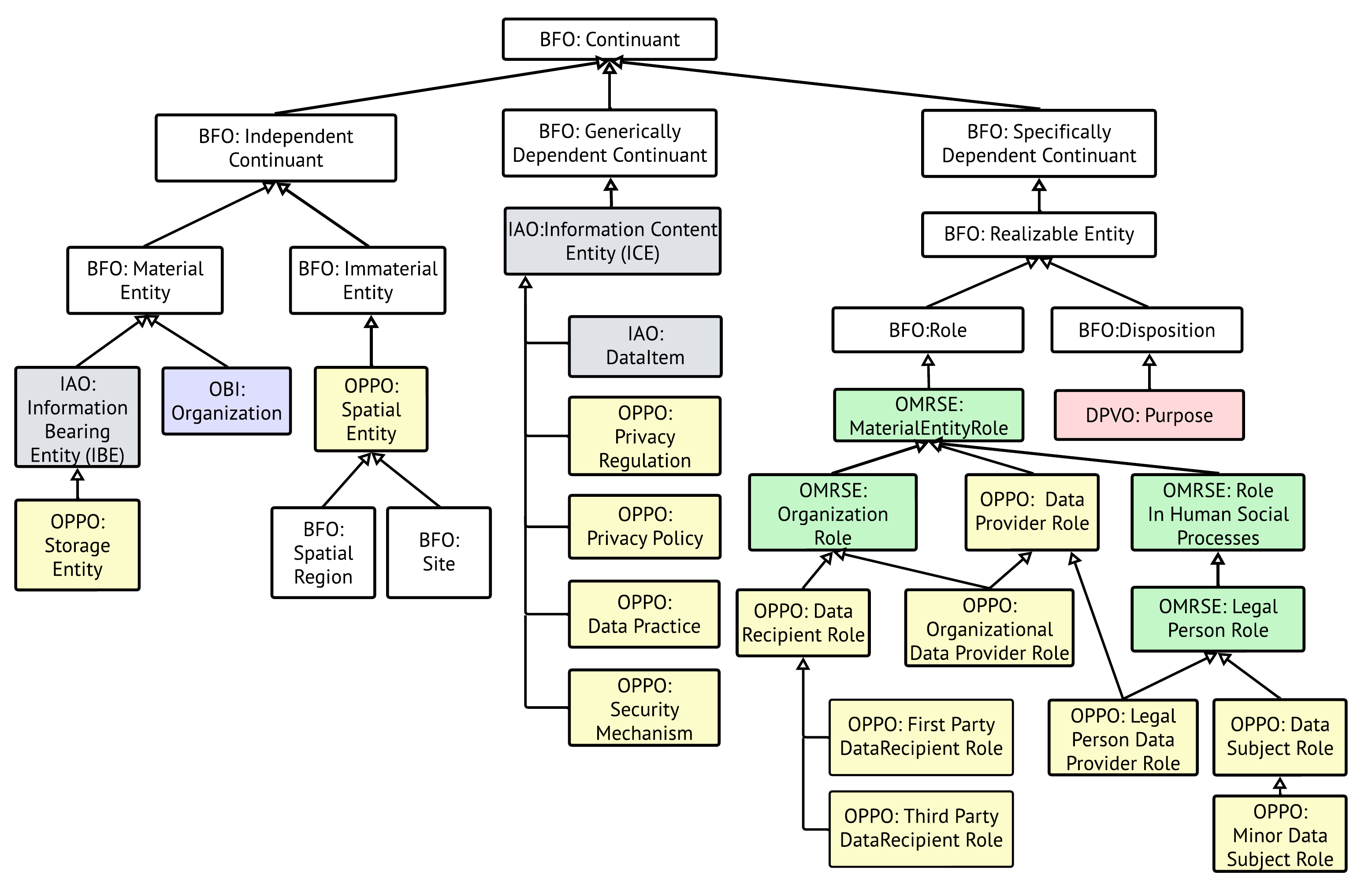}
        \caption{OPPO's high-level concepts and how they extend concepts from BFO, IAO, OBI, and OMRSE. Concepts in yellow are introduced by OPPO, while all other concepts are being reused.}
        \label{fig:OPPOCoreModule}  
\end{figure}

OPPO builds and reuses a number of existing ontologies. We directly reuse the classification of purposes from DPV \cite{pandit2019creating} and time-related concepts (e.g.\ to capture storage duration) from OWL-Time \cite{Hobbs2006} as discussed in more detail later on. In addition, we use top- and mid-level ontologies to ground OPPO concepts. Because privacy policies and data practices are primarily about information, we rely on the Information Artifact Ontology (IAO)\footnote{Available from \url{https://github.com/information-artifact-ontology/IAO}} \cite{smith2013iao} as the main source for upper-level reference concepts while  OMRSE \cite{hicks2016ontology} and OBI \cite{bandrowski2016ontology} provide additional high-level role classifications of interest to OPPO. All of IAO, OMRSE, and OBI build on the top-level ontology BFO \cite{arp2015building}. Thus, to maximize compatibility we also reuse BFO as top-level ontology. Figure~\ref{fig:OPPOCoreModule} demonstrates how OPPO's most generic concepts relate to BFO and the other ontologies. In the remainder of this section we will provide overview of OPPO's key concepts and their relationship to top- and mid-level ontologies. 

\vspace{-3pt}
\paragraph{Information content entities:}
\underline{Privacy policies} describe the \underline{data practices} an OSN employs to collect, store and process users' \underline{data} in compliance with \underline{privacy regulations}. We have identified the four underlined concepts as central to this endeavor, which all are modeled as subclasses of \texttt{IAO:InformationContentEntity}: (i) a \texttt{PrivacyRegulation} such as GDPR or CCPA that regulate the storage, collection, and processing of user's personal information; (ii) a \texttt{IAO:DataItem} is a piece of personal information directly or indirectly associated with a user (e.g., name or age); (iii) a \texttt{DataPractice} describes a way that an OSN processes user data (e.g., how long or where it retains data); and (iv) a \texttt{PrivacyPolicy} describes the entirety of the data practices of an OSN or, more generally, any kind of organization. 
Note that by modeling both \texttt{PrivacyPolicy} and \texttt{DataPractice} as subclasses of \texttt{InformationContentEntity} we make the intentional choice to treat them as the content of a privacy policy (\texttt{PrivacyPolicy})  or a portion of that content that describes a specific practice (\texttt{DataPractice}). They are distinct from the text itself and from what the OSN actually does in practice with the data.
Data practices may mention specific techniques or tools employed to protect user data, which are modeled as \texttt{SecurityMechanism}s\footnote{For brevity, we omit the \texttt{OPPO:} namespace; concepts and properties without a namespace are implicitly assumed to be within that namespace.}. 

\vspace{-3pt}
\paragraph{Roles:}
We have identified three distinct roles relevant to privacy policies and the data practice described therein, which align well with roles distinguished in the OMRSE ontology. 
(i) \texttt{DataSubjectRole} is a role played by a person whom the collected data is about (e.g.\ a user's date of birth or a message they posted). It specializes  \texttt{OMRSE:LegalPersonRole}. 
(ii) \texttt{DataRecipientRole} is a role played by an organization (\textit{OBI:Organization} in OBI's terms) that receives information either directly from a person or from a third party. It specializes \texttt{OMRSE:OrganizationRole}. 
(iii) \texttt{DataProviderRole} is a role played by either a person or an organization that shares data with others. To distinguish whether the data is shared by a person or an organization, we  introduce the subclasses \texttt{LegalPersonDataProviderRole} and \texttt{OrganizationalDataProviderRole} that also specialize \texttt{OMRSE:OrganizationRole} and \texttt{OMRSE:LegalPersonRole}, respectively.
Note also that users and organizations can play multiple roles for a particular piece of data, for example, a user can share data about themselves, in which case the user acts in both a \texttt{DataSubjectRole} and a \texttt{DataProviderRole}.  Likewise, one organization can act in a \texttt{DataRecipientRole} when it receives some data from a third party and in a \texttt{DataProviderRole} when sharing data with the same or other third parties. 

These roles can be further refined as shown in the right half of  Figure \ref{fig:OPPOCoreModule}. One additional refinement is motivated by regulations, such as GDPR and CCPA, imposing stricter conditions on handling data from minors, that is, users under a certain age (which may be defined differently by different regulations). Thus, we introduce \texttt{MinorDataSubjectRole} as a subclass of \texttt{DataSubjectRole}. 
Another distinction is between two kinds of \texttt{DataRecipientRole}s. An organization may act in a \texttt{FirstPartyDataRecipientRole} role when it receives data \textit{directly} from a person in which case the person acts in both a \texttt{DataSubjectRole} and a \texttt{DataProviderRole}. An organization may also share the collected data with other organizations according to its own data practices. Any other organization that receives such data then acts in the \texttt{ThirdPartyDataRecipientRole} and is bound not only by its own privacy policies but also by the policy of the organization that it receives the data from. For example, if an organization that is a \texttt{FirstPartyDataRecipientRole} for some piece of data says that it shares data only for specific purposes with third parties, then these third parties are expected not to share that information with others for any other purposes either.

\begin{figure}[!tb]
    \centering
   \includegraphics[width=0.7\textwidth]{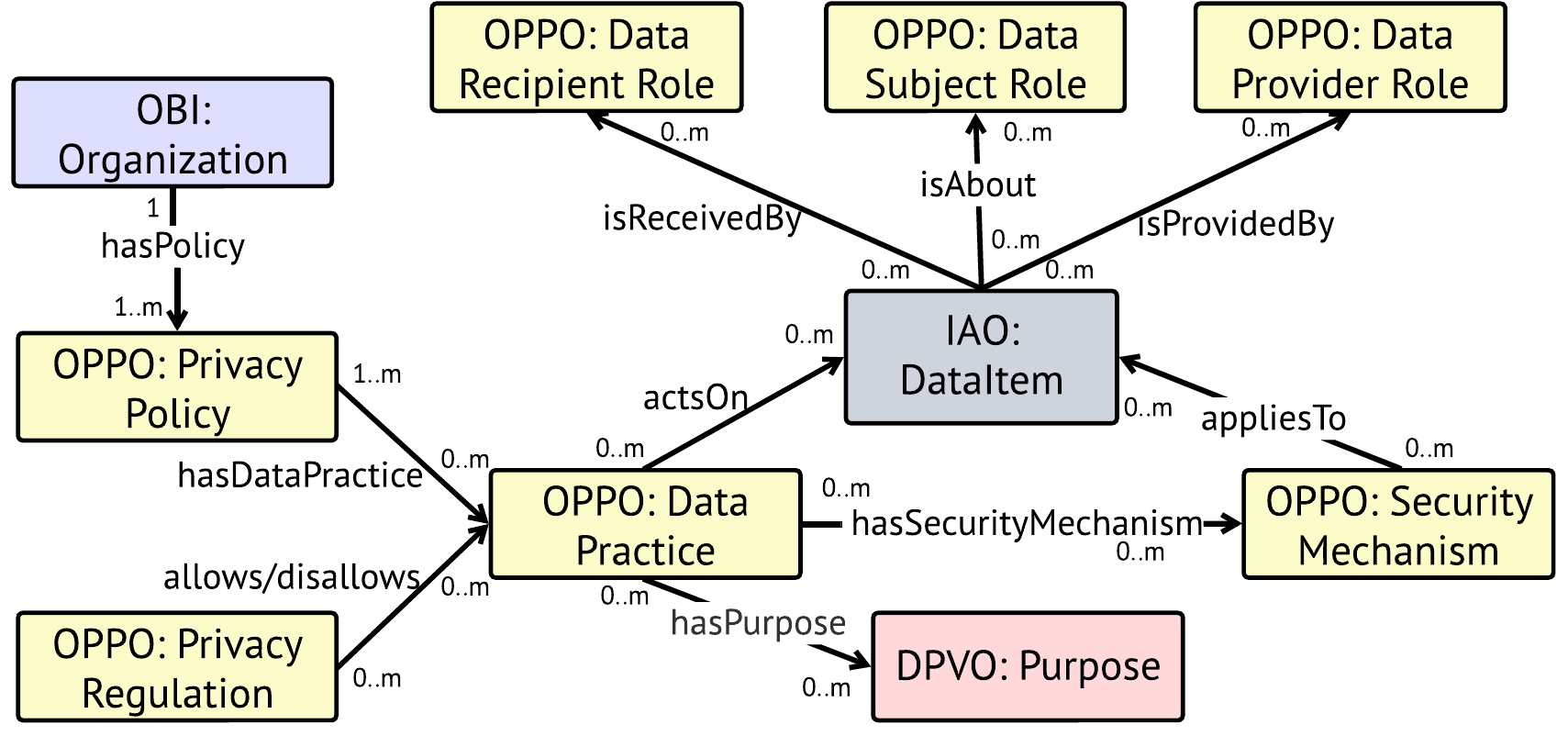}
    \caption{The core concepts in OPPO and the relations between them.}
    \label{fig:OPPOCore-ConnectionModule}
\end{figure}

\vspace{-3pt}
\paragraph{Purposes:}
Many privacy regulations, such as GDPR, require privacy policies to outline for what purpose an organization collects or processes data.  For instance, GDPR Art. 5 \cite{GDPR} states that an organization may collect different information types for different purposes such as  archiving statistical or research purpose. To model these different purposes we adopt the generic concept \texttt{DPVO:Purpose} from the DPV ontology \cite{pandit2019creating}, which is a subclass of \textit{BFO:Disposition} in BFO. Thus, all specialized purposes from DPV can be reused as well. 

\vspace{-3pt}
\paragraph{Relations among OPPO's core concepts:}
Figure \ref{fig:OPPOCore-ConnectionModule} shows the key relations among the core concepts of OPPO, forming its central Ontology Design Pattern (ODP). It consists of the following relations. 
A \texttt{PrivacyPolicy} of an OSN (modeled as an \texttt{OBI:Organization}) contains data practices ((\texttt{hasDataPractice}) relation to \texttt{DataPractice}) that describe how the OSN collects, stores, and processes  users' data. These data practices may be explicitly allowed or disallowed (\texttt{allows} and \texttt{disallows}) by different \texttt{PrivacyRegulation}. 
\texttt{DataPractice}s apply to (\texttt{actsOn}) specific kinds of data (\texttt{IAO:DataItem}), such as specific subclasses of data (e.g.\ \texttt{DemographicPersonalData} or \texttt{AnonymizedData} as elaborated in Figure~\ref{fig:datamodule}) or data that is constrained in other ways, for example, what kind of legal person the data is about (a minor or not), what purpose it is collected for, or how it was received. More specifically, \texttt{IAO:DataItem}s can be linked to specific data subjects, such as who the data is about (\texttt{isAbout} linking to a \texttt{DataSubjectRole}), data providers (\texttt{isProvidedBy} linking to a \texttt{DataProviderRole} describing who provides the data), and data recipients (\texttt{isReceivedBy} linking to a \texttt{DataRecipientRole} describing who receives the data). Because an OSN may describe different practices for storing or collecting data for distinct purposes (\texttt{DPVO:Purpose}), such as archiving or marketing purpose, a practice is related to purposes via the \texttt{hasPurpose} relation. By specifying both purposes and specific types, instances of the \texttt{DataPractice} class can capture that certain \texttt{DataItem}s (e.g.\ personal data or statistical data) are stored or processed only for certain purposes. Similarly, an OSN may employ different security mechanism in different \texttt{DataPractice} and thus, implicitly, for different kinds of \texttt{DataItem}s. For example, an organization may apply end-to-end encryption mechanism to biometric data while using a pseudonymization mechanism for personal technical data such as IP addresses. We capture this by relating \texttt{SecurityMechanism}s to \texttt{IAO:DataItem} using the \texttt{appliesTo} relation. In the next subsections, we will discuss further refinements of \texttt{IAO:DataItem}, \texttt{DataPractice}, and \texttt{SecurityMechanism}.

\subsection{Data Item Module}
\label{subsec:Data}

\begin{figure}[tb]
  \begin{center}
\centering
     \includegraphics[width=\textwidth]{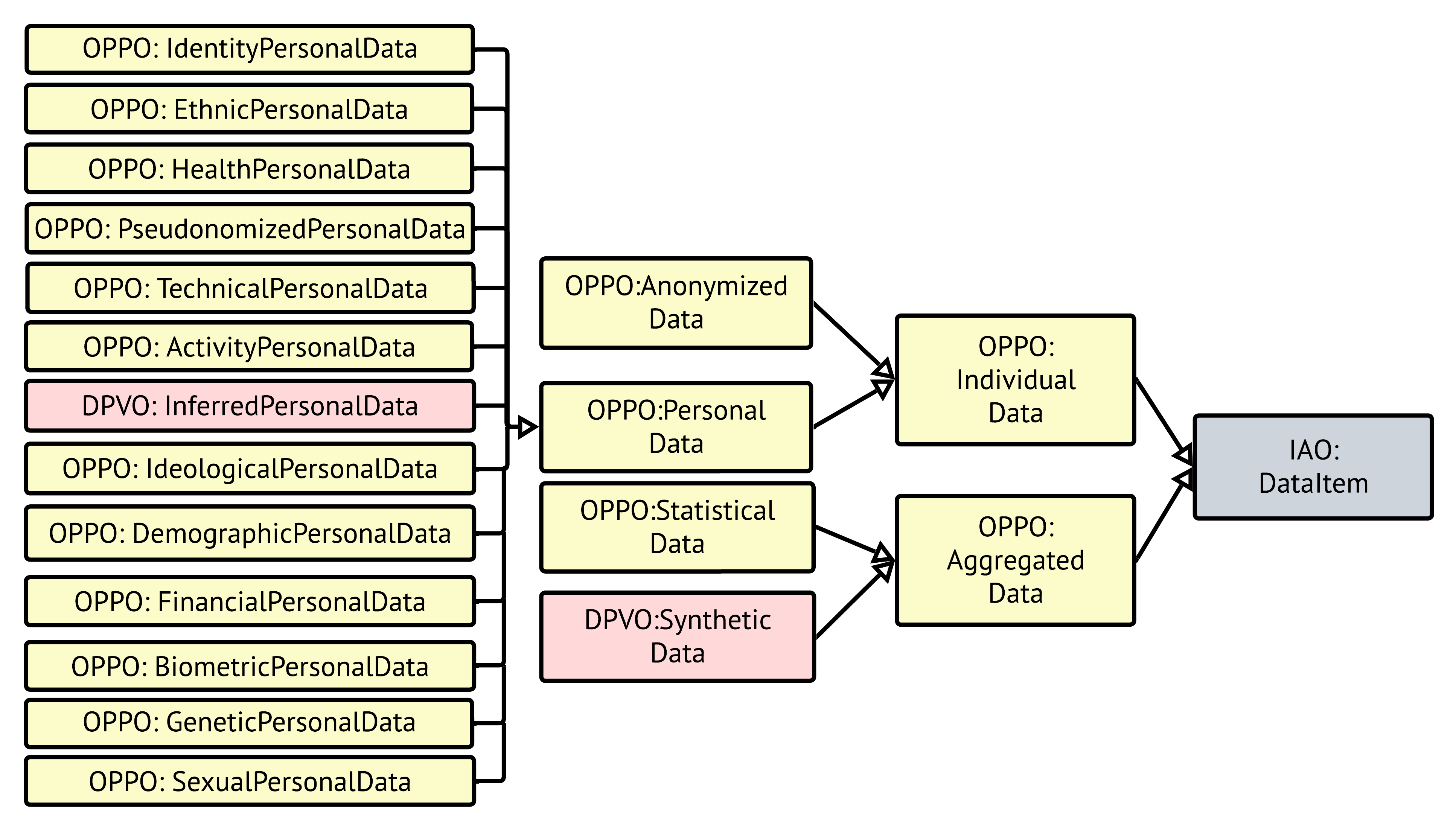}
        \caption{OPPO's taxonomy of subclasses of \texttt{IAO:DataItem}. The yellow concepts are introduced by OPPO; all other concepts are reused from DPVO and IAO.}\label{fig:datamodule}
         \end{center}
\end{figure}

To describe the different kinds of data collected by an OSN, we reuse and refine \texttt{IAO:DataItem} \cite{smith2013iao} as shown in Figure \ref{fig:datamodule}. At the highest level, we distinguish data related to an individual (\texttt{IndividualData}), such as a name, age, or credit card information, from data that is an aggregated across multiple individuals (\texttt{AggregatedData}).  
\texttt{IndividualData} can be further distinguished based on whether it is anonymous, that is, data that cannot be used to personally identify any specific individual person (\texttt{AnonymizedData}). Data which may -- directly or indirectly -- reveal an individual's identity falls into the complementary class of \texttt{PersonalData}. It includes, for example, photos, fingerprints, posts, reviews, location, or credit card information. Based on the analysis of GDPR, CCPA, privacy policies, and prior works \cite{gupta2021phin, pandit2019creating}, we have identified thirteen subclasses of \texttt{PersonalData} as shown in Figure \ref{fig:datamodule}. Their full definitions are provided in the ontology using the \texttt{skos:definition} relation. One noteworthy concept is that of \texttt{PseudonymizedPersonalData}, which is widely considered to be still \texttt{PersonalData}, though any personal identifiers have been replaced by a pseudo-identity. But it is different from \texttt{AnonymizedData} in that the data can be still ascribed to an individual by anyone who knows the mapping (or mapping algorithm) between pseudonyms and personal identifiers. 
We also include \texttt{dpvo:InferredPersonalData} from DPV as subclass of \texttt{PersonalData}. It includes any new data that is derived from existing data (e.g., demographic information from the browsing history) and which may, directly or indirectly, identify an individual. 

Finally, two distinct subclasses of \texttt{AggregatedData} are included in OPPO: (i) statistical aggregations of user data (\texttt{StatisticalData}), such as the number of views on a product page or the number of likes of a post, and (ii) artificial data produced to mimick real user data (\texttt{dpvo:SyntheticData}). 

\subsection{Data Practice Module}
\label{subsec:DataPractice}

\begin{figure}[!bt]
  \centering
     \includegraphics[width=0.90\textwidth]{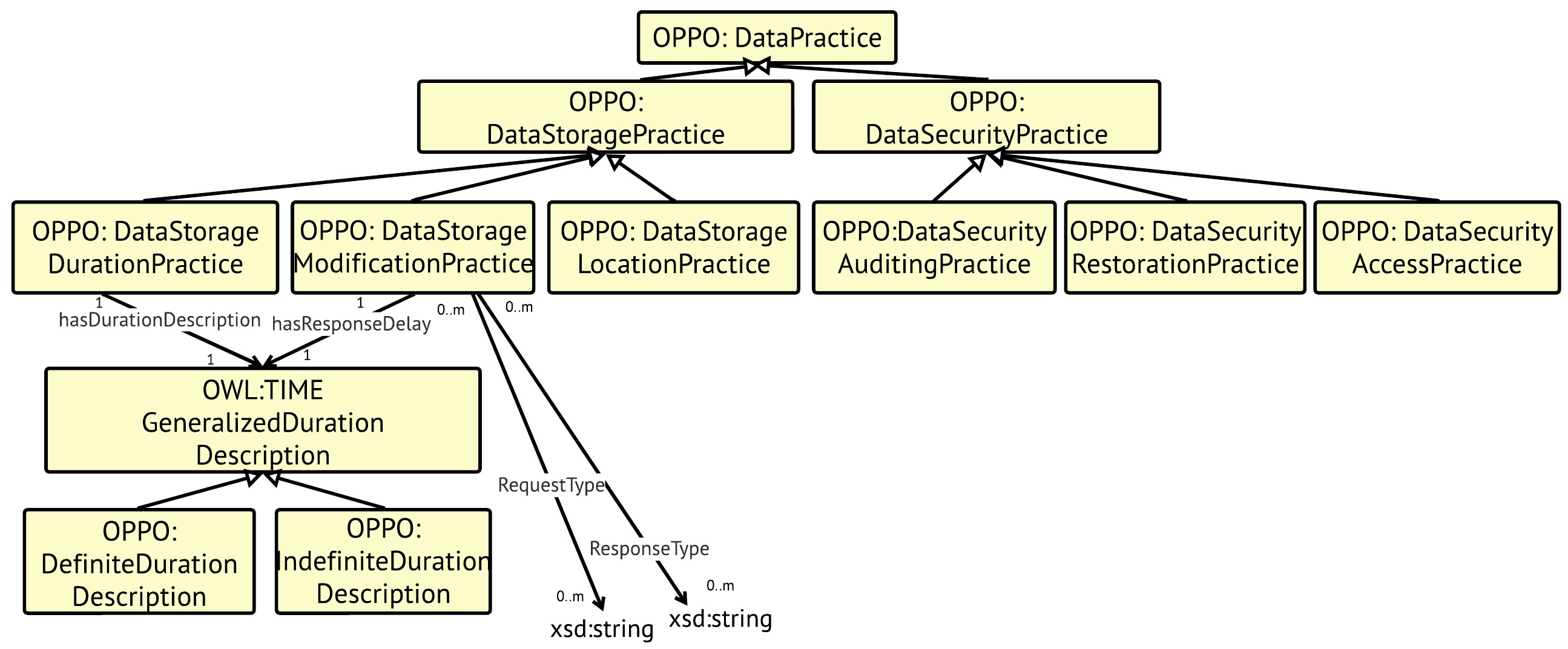}
     \caption{OPPO's taxonomy of \texttt{DataPractice}s with indicative relations for the \texttt{DataStoragePractice}s.}\label{fig:dataPracticemodule}
\end{figure}

As mentioned in Section \ref{sec:Intro}, the upper level of OPPO is currently only refined to the extent needed to model fine-grained data storage and retention  practices (\texttt{DataStoragePractice}) and security practices (\texttt{DataSecurityPractice}). For each of them, we distinguish three subclasses as shown in Figure \ref{fig:dataPracticemodule} and explained next.

\paragraph{Data Storage Practices:}
\texttt{DataStoragePractice}s can specify restrictions on the  duration of the storage, the location of the storage, and how to get stored data corrected or deleted. While all storage practices can give such details, those that do fall into a of three subclasses. A \texttt{DataStorageDurationPractice} must specify the duration of the stored data, which may be definite or indefinite. 
A \texttt{DataStorageLocationPractice} includes restrictions on where or how the data is stored. For instance, it may apply to data stored in specific geographic locations (e.g., EU-GDPR imposes restrictions on data practices while the data is stored outside of the EU) or sites of a specific company. We use the new concept \texttt{SpatialEntity} that generalizes both \texttt{BFO:SpatialRegion} and \texttt{BFO:Site} (as shown in Figure \ref{fig:OPPOCoreModule}) as location to remain  flexible and compatible with how different BFO-based ontologies may specify locations. Alternatively, a \texttt{DataStorageLocationPractice} may specify the kind of physical infrastructure (\texttt{StorageEntity}) the data is stored in, such as a data center or the user's device.

\texttt{DataStorageModificationPractice} is a subclass of \texttt{DataStoragePractice} that  specifically captures practices that allow the modification (i.e., correction or deletion) of stored data. It provides relations to specify how users can request to rectify inaccurate data (a data property \texttt{RequestType}), how the OSN may respond to such requests (a data property \texttt{ResponseType}), or how fast long the OSN may take to process such data rectification or erasure request (using \texttt{hasResponseDelay}). To capture different data retention practices as well as response delays, we reuse the \texttt{TIME:GeneralizedDurationDescription} concept and refine it by introducing \texttt{DefiniteDurationDescription} and \texttt{IndefiniteDurationDescription} as our own specializations. We refer to the ontology for details.

\paragraph{Data Security Practices:} 
\texttt{DataSecurityPractice}s are organizational data practices 
that are followed to maintain security to the collected/stored data.
OPPO distinguishes three classes of security practices (\texttt{DataSecurityPractice}). A \texttt{DataSecurityAuditingPractice} is a practice that inspects whether and how the organization maintains proper safeguard mechanisms while collecting, storing, or processing personal data. 
A \texttt{DataSecurityRestorationPractice} is a practice that discusses how data will be recovered if the data has been lost, stolen, or compromised in other ways.
A \texttt{DataSecurityAccessPractice} is a practice that limits access to the data, thus preventing unauthorized access. 

\subsection{Security Mechanism Module}
\label{subsec:SecurityMechanism}

\begin{figure}[tb]
  \begin{center}
\centering
     \includegraphics[width=\textwidth]{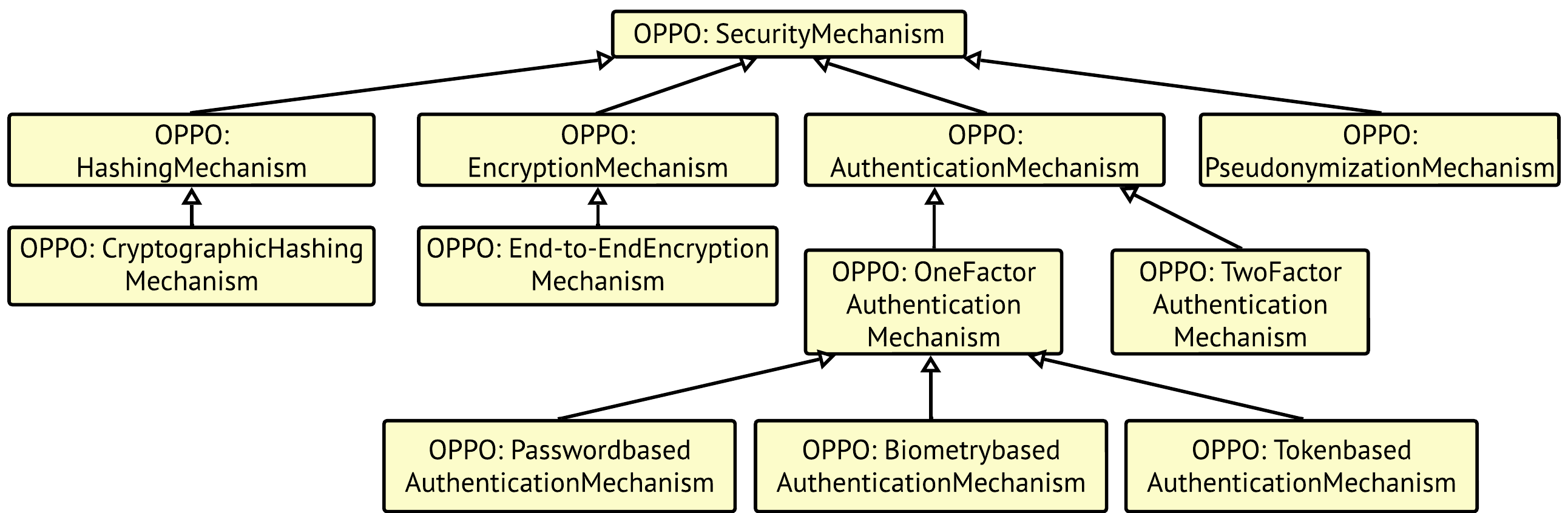}
              \caption{OPPO's taxonomy of \texttt{SecurityMechanism} and its subclasses.}\label{fig:securitymechanism}       
         \end{center}
\end{figure}

Both GDPR and CCPA require organizations to apply suitable \textit{techniques or tools} (referred to as \texttt{SecurityMechanism} in OPPO) to ensure the security of the collected data. The regulations themselves distinguish two types of security mechanisms: (i) \texttt{PseudonymizationMechanism}s that replace personal identifiers with a pseudo identity; and (ii) \texttt{EncryptionMechanism}s that make personal data unintelligible without the necessary keys for decryption. Our analysis of the ten OSN privacy policies, however, identified additional security mechanisms that are employed by these social networks. For instance, Signal's  privacy policy \cite{Signal} states that it applies cryptographic hashing mechanisms to collected data before transmitting it to their server. As another example, Telegram's policy \cite{Telegram} explicitly states that it employs (if the user enables it) two-factor authentication mechanisms to limit unauthorized access to their data. As a result, OPPO also distinguishes \texttt{HashingMechanism} and \texttt{AuthenticationMechanism} as two additional subclasses of security mechanisms, with further subclasses for \texttt{AuthenticationMechanism}.  The hierarchy of subclasses of \texttt{SecurityMechanism} is shown in Figure \ref{fig:securitymechanism}.
 
\section{Formalization and Evaluation}
\label{sec:evaluation}
The ontology is encoded using the Web Ontology Language (OWL2) as computer-interpretable format. The axiomatization currently contains  60 new classes,  17 object properties, and 271 logical axioms\footnote{The numbers are based on a core version of the ontology that only replicates some upper level BFO, IAO and OMRSE concepts, but those have been excluded in the concept and property counts.}
The syntactic correctness of the ontology has been verified with a simple RDFS validator\footnote{\url{http://rdfvalidator.mybluemix.net/}}. We further checked for common pitfalls in the ontology such as missing domain and range restrictions using the OntOlogy Pitfall Scanner (OOPS!) \cite{poveda2014oops}. 

\vspace{-3pt}
\paragraph{Verification:}
\label{subsec:LogicalConsistency}
We used the HermiT  \cite{glimm2014hermit} OWL2 reasoner that is provided with Protégé to check the ontology for logical consistency. In the current version, no inconsistencies are found nor are any classes inferred as being  equivalent in any problematic ways, such as being equivalent to \texttt{owl:Thing}. 
Additionally, we manually created instances of the classes and properties from Telegram's real privacy policy\footnote{The dataset can be found in our GitHub repository.}. When re-checked the consistency of this dataset together with the ontology using the HermiT reasoner. 

\begin{figure*}[!tp]
\centering
\includegraphics[width=\textwidth]{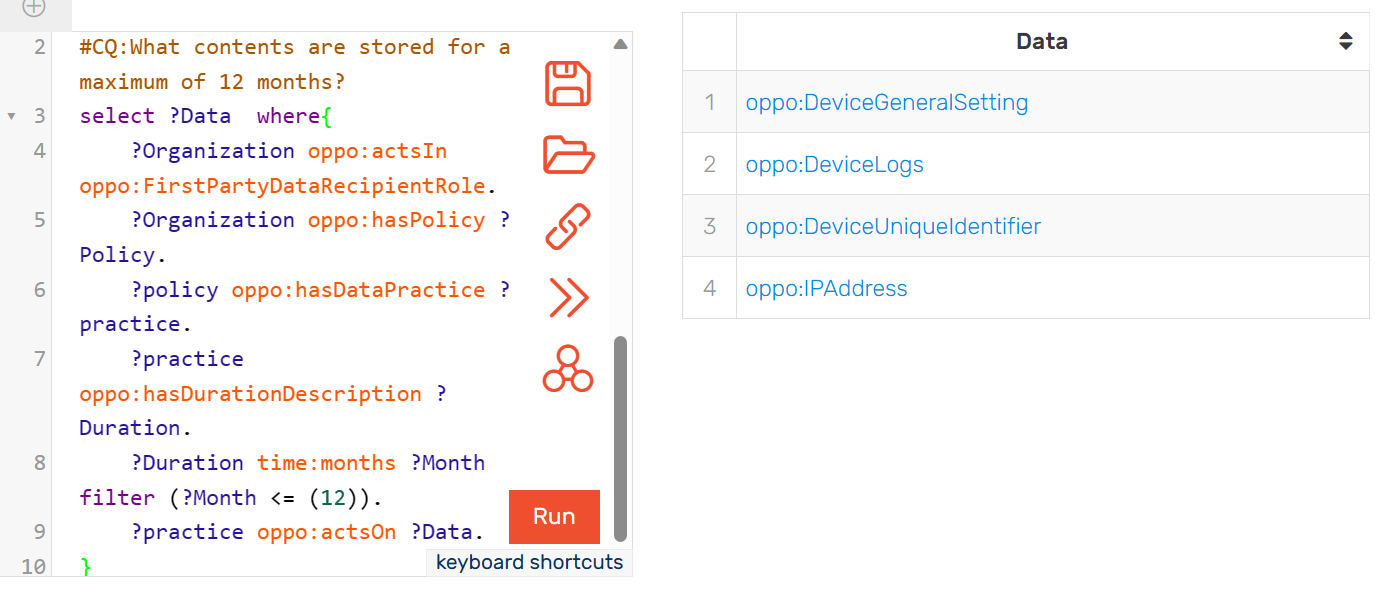}
\caption{Encoding of a sample competency question as a SPARQL Query and the answer when executed within GraphDB over the ontology and test dataset.}
\label{fig:CQ}
\end{figure*}

\vspace{-3pt}
\paragraph{Validation:}
\label{subsec:Expressivity}
To test the expressivity of OPPO, we were able to express 15 out of a total of 27 competency questions as SPARQL queries\footnote{The SPARQL queries are provided in our GitHub Repository.}. The remaining questions will guide the refinement and further development of the ontology. 
We loaded the ontology and the sample dataset into a GraphDB instance \cite{GraphDB} that supports RDF and OWL reasoning. We selected the OWL-RL (Optimized) ruleset provided by GraphDB that implements OWL2 reasoning with the limitations described by the OWL2 RL profile. We executed each SPARQL query and analyzed the results to ensure that they match what we expect for our sample data. This validated that the ontology is sufficiently expressive to adequately encode and answer these competency questions. For instance, Figure \ref{fig:CQ} shows one example of a CQ that focuses on capturing the \textit{specific data types} that are being stored by Telegram for a\textit{ maximum of 12 months}. The output indicates that Telegram  stores four types of data for a maximum of 12 months. A more in-depth evaluation of the CQs on a larger dataset will be completed in the near future.

\section{Conclusion and Future Work}
\label{sec:Conclusion_Future}
In this paper, we presented OOPO as an extensible ontology for the privacy domain that is designed to model and formally encode detailed data practices as described in OSNs' privacy policies. As a proof-of-concept, the ontology provides a core pattern that connects privacy policies and their contained data practices to data items that are described by their data types and roles. The pattern identifies refined  data practices based on the kinds of constrains they impose (e.g. the duration, location, or type of practice) in order to allow formally representing and reasoning over data practices of different privacy policies of online social networks and similar companies. The ontology leverages and connects to a number of existing ontologies based on the Basic Formal Ontology (BFO). OPPO is encoded in OWL2 and provided as an open source resource to the community. 
We evaluated the ontology's logical consistency with some small but real dataset and using a set of competency questions. The  competency questions were encoded in SPARQL, with the answers validating that the ontology can indeed express and answer queries about the details of data practices from privacy policies. 

In the future, we plan to improve OPPO by (i) extending the ontology with other practices, such as those related to data collection and sharing;  by (ii) constructing larger datasets from multiple privacy policies in order to evaluate and compare their degrees of transparencies while also further evaluating the ontology; and by (iii) expanding the ontology to also model different privacy regulation in order to pose and answer questions about compliance and non-compliance between individual privacy policies and different privacy regulations, which is becoming increasingly important as more regulations emerge.

\vspace{-3pt}
\paragraph{Acknowledgments}
We thanks the anonymous reviewers for the helpful suggestions to improve the final version of the paper. 

\bibliography{oppo-bib}

\end{document}